\documentclass[%
reprint,
 amsmath,amssymb,
 aps, physrev,
]{revtex4-2}
\pdfoutput=1

\usepackage{graphicx}% Include figure files
\usepackage{dcolumn}% Align table columns on decimal point
\usepackage{bm}% bold math
\usepackage{xcolor}
\usepackage{longtable}

\graphicspath{{./}{figures/}}

\begin{document}

\title{Sibling Rivalry: Thermonuclear Diversity and the Hubble Tension}% Force line breaks with \\

\author{Richard S. Miller}
 \email{Richard.S.Miller@jhuapl.edu}
\affiliation{%
 Johns Hopkins University Applied Physics Laboratory \\
11100 Johns Hopkins Road, Laurel, MD 20723, USA \\
}

\date{\today}% It is always \today, today,

\begin{abstract}
Homogeneity is the hallmark of standard candle-based cosmology investigations. Thermonuclear supernovae (Type-Ia, SNeIa) violate this essential requirement if they develop along multiple evolutionary pathways. In this work, the impact of thermonuclear diversity on cosmological parameter constraints is quantified using Pantheon+, one of the largest ensembles of SNeIa compiled to probe cosmology to date. Evidence of diversity is encoded in supernova light curves. Pantheon+ is shown to be diverse, with features indicative of multiple thermonuclear sub-classes. Diversity-driven systematic effects have been quantified on a supernova-by-supernova basis; event selections based on light curve-derived metrics were subsequently used to characterize diversity-dependent trends and mitigate their impact. A "diversity free" estimate of the Hubble-Lema\^{i}tre parameter, $H_0$=$67.5\pm3.5$ km s$^{-1}$ Mpc$^{-1}$ (68\% C.L.), was obtained by reanalyzing Pantheon+. The Hubble Tension, an apparent disparity between early- and late-Universe determinations of $H_0$, is eased from $\sim$$5\sigma$ to $<$$1\sigma$ after accounting for the diverse thermonuclear scenarios that govern SNeIa. A strategy for precise determination of $H_0$, dominated by statistical rather than systematic uncertainties, is also presented.

%\begin{description}
%\end{description}
\end{abstract}

\maketitle

%\tableofcontents

\section{\label{sec:intro}Introduction}

Thermonuclear supernovae (Type-Ia, SNeIa) are one of the premiere probes of cosmology. Their treatment as standardizable candles and visibility out to high redshifts has had sweeping ramifications, including direct evidence for an accelerating universal expansion and the existence of dark energy \cite{1998AJ....116.1009R, 1999ApJ...517..565P}. Type-Ia supernovae are witnesses to both cosmic and stellar evolution, and continue to be at the forefront of cosmology investigations.

Supernovae probe the history of cosmic expansion to constrain cosmological models. Accurate constraints require a Hubble Diagram defined by homogeneous cosmic distance indicators, i.e., standard (or standardizable) candles \citep{2011ARNPS..61..251G}. Non-standard candles mask the true record of expansion and can bias model interpretation. 

The essential role of SNeIa as distance indicators belies the fact that they are an enigma. Type-Ia supernovae may develop along multiple evolutionary pathways: Rapid thermonuclear burning of a white dwarf is central to the SNeIa paradigm \cite{1984ApJ...286..644N, 2006NuPhA.777..579H}, but their progenitor(s), method(s) of ignition, and degree of intrinsic heterogeneity remain a topic of intense debate \cite{2000ARA&A..38..191H,2014ARA&A..52..107M,NAP12951,NAP26141}. Diversity is the upshot of a SNeIa population defined by a range of thermonuclear scenarios.

Cosmology investigations rely on optically-detected SNeIa; event selections utilize criteria based on indirect secondary signatures rather than their thermonuclear identities \cite{1993ApJ...413L.105P,2007A&A...466...11G}. As a result, the homogeneity of supernova datasets cannot be guaranteed. Systematic effects are a natural consequence of event ensembles selected from a diverse population.

Advances in understanding the sources of systematic error are needed to address open questions in cosmology and ensure the most accurate model constraints. Thermonuclear diversity has been overlooked in this regard; its impact on cosmological model constraints has not been comprehensively evaluated. Diversity-driven systematic effects, and their impact, are quantified here using Pantheon+, one of the largest SNeIa ensembles compiled for cosmological analyses to date. 

The imperative for a deeper understanding of systematic effects is motivated in part by the Hubble Tension, an unresolved disparity in measurements of the Hubble-Lema\^{i}tre parameter ($H_0$), the fundamental metric of cosmic expansion \citep{2021ApJ...919...16F,2021CQGra..38o3001D,2022arXiv221104492K,2022JHEAp..34...49A,2022NewAR..9501659P}. The apparent $\sim$$5\sigma$ inconsistency between early- (redshifts $z\geq1100$) and late-time ($z\sim\mathcal{O}(1)$) determinations of $H_0$ is one of the key outstanding mysteries of modern cosmology \cite{2020A&A...641A...6P,2022ApJ...938..110B}. SNeIa are integral to the late-time estimates.

This paper is organized as follows: In Section~\ref{sec:lightcurves}, the use of light curves as a probe of thermonuclear diversity is motivated. In Section~\ref{sec:diversity_metric}, diversity-driven systematic effects are quantified; evidence of their origin is isolated using supernova siblings. In Section~\ref{sec:cosmology}, the framework for obtaining cosmological constraints is reviewed. In Section~\ref{sec:impact}, the impact of thermonuclear diversity on cosmological constraints is quantified, and in Section~\ref{sec:tension}, the status of the Hubble Tension is updated. Key takeaways are summarized in Section~\ref{sec:summary}, along with a path forward for future supernova cosmology investigations.

\begin{figure}[!]
    \includegraphics*[scale=0.25]{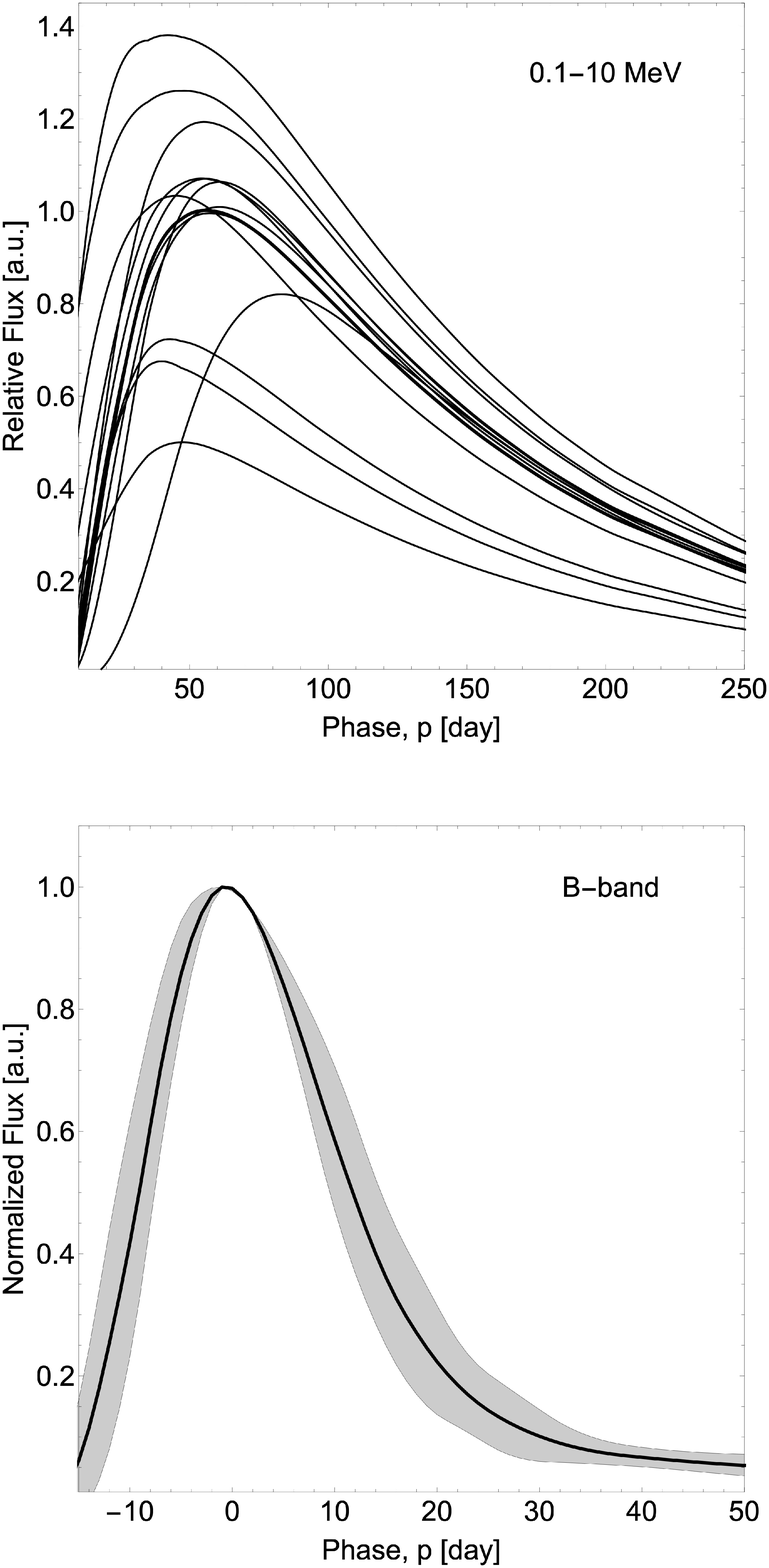}
    \caption{\label{fig:shape_examples}SNeIa light curves. (top) Emergent nuclear gamma rays (0.1-10 MeV) from representative SNeIa models \cite{2014ApJ...786..141T}. (bottom) Standardized Pantheon+ $B$-band light curves. The average SALT2 light curve ($x_1$=$c$=0, black) and range of shapes (gray) are shown. Peak $B$-band emission is engineered to occur at $p$=0 per SALT2; shape differences have been isolated by normalizing the peak to unity. Uncertainties have been removed for clarity.} 
\end{figure}

\section{\label{sec:lightcurves}Light Curves as a Proxy \\ for (Thermonuclear) Diversity} 

Clues to the thermonuclear identity of SNeIa are encoded in their light curves, the temporal evolution of emergent radiation. The shape of each light curve is governed at a fundamental level by details of the explosion process, the mass and volumetric distribution of $^{56}$Ni, internal structure and composition, as well as the time-dependent ejecta opacities \cite{2017hsn..book.1955S,1969ApJ...155...75C,2008MNRAS.385.1681S,2014ApJ...786..141T,2021MNRAS.508.1590P,2013A&A...554A..67S}.

As nuclear radiation from the $^{56}$Ni$\rightarrow$$^{56}$Co$\rightarrow$$^{56}$Fe decay chain diffuses through the ejecta, adiabatic losses heat the optically thick material and power its expansion. Peak emission occurs once the ejecta becomes optically thin, i.e., when the age of the ejecta exceeds the (energy-dependent) diffusion timescale. Optical emission, the foundation of supernova cosmology, is a by-product of the decay-diffusion-expansion process.

Emergent nuclear radiation (gamma rays) is the most direct probe of SNeIa thermonuclear scenarios. Nuclear light curves have relatively simple single-peaked shapes, independent of the underlying physical model (e.g., Fig.~\ref{fig:shape_examples}). Thermonuclear sub-classes can, in theory, be differentiated by characterizing light curve shapes (e.g., \cite{2008MNRAS.385.1681S,2013A&A...554A..67S}), but gamma rays from just one SNeIa have been observed to date \cite{2014Sci...345.1162D,2014Natur.512..406C,2015ApJ...812...62C,2016A&A...588A..67I}. The degree of thermonuclear diversity remains unknown.

Emergent optical radiation affords deep surveys and large numbers of detected events. The cost is information loss, an adverse effect of the nuclear radiation reprocessing that gives rise to the secondary emission \citep{1969ApJ...157..623C}. Complex interstellar and circumstellar environments also affect the optical signatures, independent of intrinsic supernova properties \cite{2017ApJ...842...93M,2021MNRAS.507.4367C}. Connections (if any) between secondary optical signatures and specific thermonuclear scenarios have not been established. 

Peak optical emission is a staple of supernova cosmology investigations (Sec.~\ref{sec:cosmology}). ``Arnett’s Rule'' states that the luminosity at maximum light is equal to the instantaneous rate of energy deposition from the $^{56}$Ni decay chain to the expanding optically thin ejecta \cite{1979ApJ...230L..37A,1982ApJ...253..785A,1985Natur.314..337A}. However, the accuracy of this rule is an open question \cite{1993A&A...270..223K,2006A&A...460..793S,2013MNRAS.429.2127B,2014MNRAS.440.1498S,2016A&A...588A..84D,2017ApJ...846...58H,2019ApJ...874...62S}, and it may only be valid for SNeIa with well-mixed $^{56}$Ni concentrations \cite{2019ApJ...878...56K}. 

Peak luminosity that depends on the thermonuclear characteristics of each supernova must also be dependent on light curve shape. Simple single-peaked light curves can be ``standardized'' to conform to a reference shape by being stretched or squeezed, but this must be accompanied by adjustments in peak amplitude to conserve flux. The distribution of shape-dependent amplitude adjustments is a proxy for thermonuclear diversity.

\subsection{\label{sec:pantheon_LCs}Pantheon+ $B$-band light curves} 

Pantheon+ consists of 1701 light curves from 1550 distinct SNeIa that span redshifts in the range $0.001\leq z\leq2.26$ \cite{2022ApJ...938..111B,2022ApJ...938..113S}. The light curves were compiled from 18 different surveys, not all of which utilize the same photometric system \cite{2022ApJ...938..113S}. Each member of Pantheon+ was standardized using the empirical Spectral Adaptive Light-curve Template (SALT2) model \cite{2007A&A...466...11G}.   

Standardization returns three parameters for each dataset member: an amplitude ($x_0$), light curve stretch ($x_1$), and color ($c$). To ensure a uniform analysis across the dataset, standardized $B$-band light curves were constructed by integrating the SALT2 emergent flux model,
\begin{equation}\label{eqn:salt2flux}
F\left(p,\lambda\right)=x_0\left[M_0\left(p,\lambda\right)+x_1M_1\left(p,\lambda\right)\right]e^{cCL\left(\lambda\right)},
\end{equation}
over a bandwidth $\Delta\lambda$=94 nm (effective midpoint $\lambda_{\mathrm{eff}}$=445 nm),
where $p$ (phase) is the rest-frame time since maximum $B$-band luminosity and $\lambda$ is the rest-frame emission wavelength; $M_0\left(p,\lambda\right)$ and $M_1\left(p,\lambda\right)$ describe the average SNeIa spectral energy distribution (SED) and its main variability components \cite{2007A&A...466...11G,2022ApJ...938..111B}, respectively, and $CL\left(\lambda\right)$ is the average color correction law \cite{2007A&A...466...11G}.

All data products required for light curve construction are available from the Pantheon+ public archive \cite{2022ApJ...938..110B}. Archived SEDs have wavelength and phase resolutions of 0.5 nm (200-900 nm) and 1 day, respectively. Flux uncertainties at each wavelength and phase were estimated using the partial derivatives of Equation~\ref{eqn:salt2flux} and archived standardization parameter uncertainties. 

Standardized light curves should be identical, but differences endure if the process is incomplete. The distribution of standardized Pantheon+ $B$-band light curves is shown in Fig.~\ref{fig:shape_examples}. Their simple single-peaked shapes echo the general features of nuclear light curves; the assortment of shapes is qualitatively inconsistent with the standard candle hypothesis. 

\begin{figure}[!]
    \includegraphics[scale=0.375]{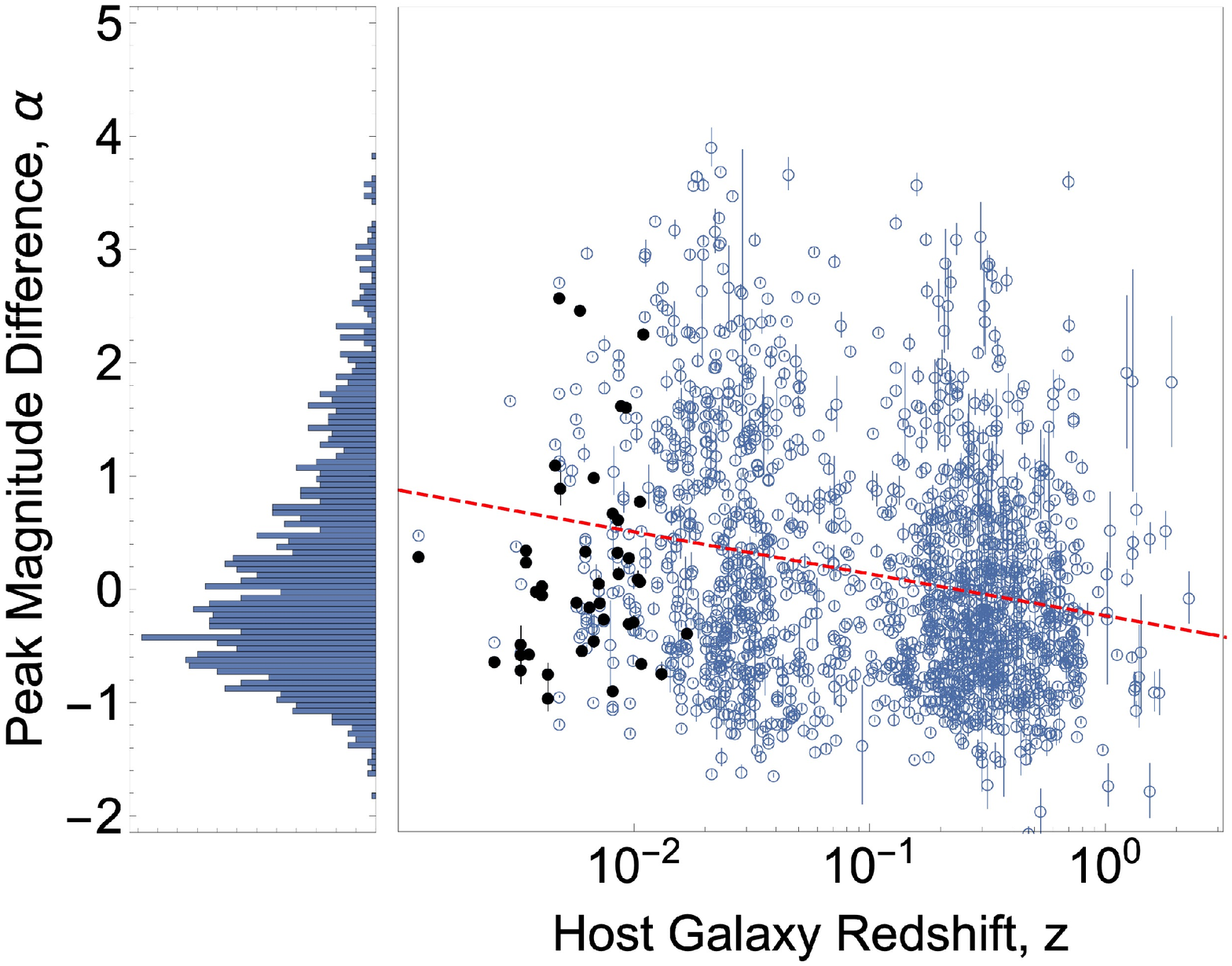}
    \caption{\label{fig:sys_vs_redshift}Pantheon+ peak $B$-band magnitude differences ($\alpha$). Shape-dependent differences in peak magnitude relative to the average SALT2 $B$-band light curve ($x_1$=$c$=0). The relative $\alpha$ value defined by Equation~\ref{eqn:magnitude_difference} (with $A$=$A_{\mathrm{ref}}$) is shown for each member of the Pantheon+ sample. The calibrator SNeIa used to anchor host galaxy distances are also shown (black). The redshift dependence is illustrated with a fit of the form $\Delta\left(z\right)=a+b\log_{10}\left(z\right)$; it is provided only to highlight a systematic trend of decreasing $\alpha$ with increasing redshift  (red, dashed).} 
\end{figure}

\section{\label{sec:diversity_metric}Quantifying Diversity} 

In order to quantify diversity within Pantheon+, each $B$-band light curve was fit using a mathematical description agnostic to any specific thermonuclear scenario. The parameterization consisted of two quasi-exponential processes, one that represents energy injection into the system, and another its removal \cite{2005ApJ...627..324N,2012MNRAS.419.1650N},
\begin{equation}\label{eqn:lcpt}
I\left(p\right)=Ae^{-\xi\left(\tau/p+p/\tau\right)}.
\end{equation}
Fit parameters have nearly orthogonal meanings: $A$ (flux normalization) scales the amplitude only; $\tau$ (epoch of peak emission) stretches the phase axis only; $\xi$ (asymmetry) modifies the scale-independent shape and amplitude. 

Light curve fits were restricted to a phase range that encompasses the best sampled post-explosion epochs of most SNeIa ($-5\leq p\leq20$). Fit quality was good for all Pantheon+ members (mean $\chi^2$ $p$-value of 0.99). 

Shape-dependent peak fluxes, $I\left(\tau\right)$, were used to characterize the dispersion of standardized light curve shapes. Peak flux was quantified relative to a common reference via the ratio,
\begin{eqnarray}\label{eqn:flux_ratio}
P/P_{\mathrm{ref}}&=&I\left(\tau\right)/I_\mathrm{ref}\left(\tau_\mathrm{ref}\right) \\
    &=&\left({A/A_{\mathrm{ref}}}\right)e^{-2\left(\xi-\xi_{\mathrm{ref}}\right)}, \nonumber
\end{eqnarray}
where the parameters ($A_{\mathrm{ref}}, \tau_{\mathrm{ref}}, \xi_{\mathrm{ref}}$) describe a reference light curve. The corresponding difference in peak magnitude is
\begin{eqnarray}\label{eqn:magnitude_difference}
    \alpha&=&2.5\log_{10}\left(P/P_{\mathrm{ref}}\right) \\
    &=&2.5\log_{10}(A/A_{\mathrm{ref}})-5\left(\xi-\xi_{\mathrm{ref}}\right)/\log\left(10\right), \nonumber
\end{eqnarray}
where, $\alpha$$>$0 ($<$0) indicates a flux excess (deficit) relative to the reference. Equation~\ref{eqn:magnitude_difference} describes a shape-dependent modification to standardized peak magnitudes; it is impractical since the $A_{\mathrm{ref}}$ for each host galaxy cannot be determined independently. A relative measure of peak magnitude differences is obtained by setting $A$=$A_{\mathrm{ref}}$ to isolate the shape dependence. Standard candles are expected to have a single-valued $\alpha$-distribution.

The $\alpha$-distribution for standardized Pantheon+ $B$-band light curves is shown in Fig.~\ref{fig:sys_vs_redshift} for $\xi_{\mathrm{ref}}$=2.74, a reference that describes the average SALT2 $B$-band light curve ($x_1$=$c$=0). An alternative reference shifts, but does not change, the distribution. The distribution's non-zero variance ($\sigma_\alpha^2$=0.966 mag$^2$) is indicative of shape-dependent diversity, i.e., a collection of non-standard candles.

\begin{table*}
\caption{\label{tab:siblings_table}Supernova Siblings Analysis Summary. Natural siblings control for both source distance and global host galaxy properties; adopted siblings control for distance only.}
\begin{ruledtabular}
\begin{tabular}{ccccccc}
Sibling & 
$N_p$\footnote[1]{Number of sibling pairs.} & $\overline{z}$\footnote[2]{Mean redshift; host galaxy redshifts from \cite{2022ApJ...938..110B,2020ApJ...896L..13S,2022ApJ...938..113S}.} & 
$\sigma_{\Delta\alpha}^2/\sigma_\mathrm{boot}^2$\footnote[3]{$F$-test significance in parentheses.} &
$\alpha_\mathrm{sib}$\footnote[4]{Mean and standard error; $t$-test significance in parentheses.} & 
$\Delta\mu$\footnotemark[4] &
$\Delta\mu-\alpha_\mathrm{sib}$\footnotemark[4] \\ \hline
\multicolumn{7} {c}{\it{Natural Siblings}} \\ \hline 
Pantheon+ & 23 & 0.011 & 0.467 ($2.2\sigma$)& $0.242\pm0.042$ ($4.5\sigma$)& $0.212\pm0.035$ ($4.6\sigma$)& $-0.030\pm0.044$ ($0.7\sigma$)\\ 
DES & 8 & 0.480 & 2.217 ($0.6\sigma$)& $0.578\pm0.215$ ($2.2\sigma$)& $0.343\pm0.057$ ($3.5\sigma$)& $-0.235\pm0.215$ ($1.0\sigma$)\\ \hline 
\multicolumn{7} {c}{\it{Adopted Siblings}} \\ \hline 
Pantheon+ & 45 & 0.011 & 1.194 ($2.1\sigma$)& $0.414\pm0.045$ ($6.8\sigma$)& $0.255\pm0.026$ ($7.0\sigma$)& $-0.159\pm0.035$ ($4.1\sigma$)\\
 & 13 & 0.127 & 0.868 ($1.6\sigma$)& $0.197\pm0.040$ ($3.6\sigma$)& $0.146\pm0.040$ ($3.0\sigma$)& $-0.051\pm0.064$ ($0.8\sigma$)\\
 & 13 & 0.480 & 0.704 ($0.6\sigma$)& $0.229\pm0.048$ ($3.5\sigma$)& $0.130\pm0.039$ ($2.7\sigma$)& $-0.099\pm0.050$ ($1.8\sigma$)\\
 & 1 & 1.454 & - & 0.379 & 0.129 & -0.250  
\end{tabular}
\end{ruledtabular}
\end{table*}

\subsection{\label{sec:evidence-siblings}Supernova Siblings}

The plausible origin of shape-dependent diversity was identified using "supernova siblings", i.e., SNeIa known to have the same cosmic reach (i.e., redshift). Siblings afford unique experimental controls that isolate intrinsic differences from extrinsic influences. 

Two classes of siblings were considered: 1) “Natural” siblings originate from the same host galaxy to control for both source distance and global host galaxy properties, and 2) “adopted” siblings in different host galaxies with (near-)identical redshifts that control only for distance. 

A total of thirty-one (31) natural sibling pairs were previously identified: twenty-three ($N_p$=23) in Pantheon+ (0.004$\leq$$z$$\leq$0.025) \citep{2022ApJ...938..113S} and eight ($N_p$=8) within the Dark Energy Survey (DES) sample (0.228$\leq$$z$$\leq$ 0.648) \citep{2020ApJ...896L..13S}. Seventy two ($N_p$=72) adopted sibling pairs were independently identified within Pantheon+ using the criterion of identical Hubble diagram redshifts ($z$$\geq$0.004). Siblings were processed as described in Sections~\ref{sec:pantheon_LCs} and ~\ref{sec:diversity_metric}. 

A bootstrap approach was used to determine if the variance of sibling $\alpha$-differences ($\sigma_{\Delta\alpha}^2$) was consistent with expectations from Pantheon+. Bootstraps consisted of $10^4$ trials of $N_p$ random Pantheon+ pairs drawn from a redshift range commensurate with each sibling dataset. A pooled variance of paired $\alpha$-differences was computed for each bootstrap ($\sigma_\mathrm{boot}^2$). The ratio $\sigma_{\Delta\alpha}^2/\sigma_\mathrm{boot}^2$ for each dataset is given in Table~\ref{tab:siblings_table}; consistency between sibling and bootstrap dispersion was established for each dataset using an $F$-test for equal variances.  

Siblings uniquely enable an absolute, rather than relative, determination of $\alpha$. For each sibling pair, $\alpha_\mathrm{sib}$(=$\lvert\alpha\rvert$) was obtained using Equation~\ref{eqn:magnitude_difference} and the substitution ($A$, $\xi$; $A_\mathrm{ref}$, $\xi_\mathrm{ref}$)$\rightarrow$($A_1$, $\xi_1$; $A_2$, $\xi_2$), where the numerical subscripts denote individual siblings; positive values were adopted since the ordering of siblings is arbitrary. The mean and standard error of $\alpha_\mathrm{sib}$ for each sibling dataset are given in Table~\ref{tab:siblings_table}; a siblings-as-twins hypothesis ($\alpha_\mathrm{sib}$=0) was rejected for each dataset with a high degree of confidence.

Shape-dependent differences between sibling pairs are significant, and their dispersion is consistent with Pantheon+. Together, the sibling-based statistical analyses provide strong evidence that the diverse standardized $B$-band light curves of Pantheon+ (Fig.~\ref{fig:sys_vs_redshift}) are the result of intrinsic SNeIa properties (i.e., thermonuclear diversity) rather than external factors.

\section{\label{sec:cosmology}Cosmology}

The impact of intrinsic diversity on cosmological model constraints requires a framework for obtaining those constraints. The present work employs the same formalism used previously by Pantheon+ \cite{2022ApJ...938..110B}; it is reviewed here for completeness and to provide context. 

Parameter constraints are obtained by minimizing a $\chi^2$ likelihood \citep{1998AJ....116.1009R,2006A&A...447...31A,2011ApJS..192....1C,2022ApJ...938..110B,2024arXiv240102929D}. Each likelihood is defined by  
\begin{equation}\label{eqn:likelihood}
    -2\log{\cal{L}}=\chi^2=\Delta\bm{D}^T\,C^{-1}_{\mathrm{stat+sys}}\,\Delta\bm{D}
\end{equation}
where $\Delta\bm{D}$ is a vector of SNeIa distance-modulus residuals and $C_{\mathrm{stat+sys}}$ is a covariance matrix that includes both statistical and systematic uncertainties. The residuals are defined by
\begin{equation}\label{eqn:residuals2}
    \Delta\bm{D}_i=
    \begin{cases}
        \mu_i-\mu^{\mathrm{cal}}_i,\,i\in\mathrm{host\,galaxy\,calibrator} \\
        \mu_i-\mu_{\mathrm{model}}(z_i),\,\mathrm{otherwise},
    \end{cases}
\end{equation}
where $\mu_i$, $\mu^{\mathrm{cal}}_i$, and $\mu_{\mathrm{model}}(z_i)$ are inferred, calibrator, and predicted distances, respectively. 

Inferred distances are determined using the observed properties of detected supernovae. They combine standardization parameters (Sec.~\ref{sec:pantheon_LCs}) with a modified version of the empirical Tripp relation \cite{1998A&A...331..815T,2022ApJ...938..110B},
\begin{equation}\label{eqn:tripp}
\mu_i=m_{B,i}+\alpha x_{1,i}-\beta c_i-M
\end{equation}
where, $\alpha$ and $\beta$ are correlation coefficients of luminosity with $x_1$ and $c$, respectively \cite{2007A&A...466...11G,2017ApJ...842...93M,2022ApJ...938..113S}, $M$ is the fiducial absolute magnitude of SNeIa, and peak $B$-band magnitude is given by $m_B=-2.5\log_{10}\left(x_0\right)$. Phenomenological “broader-brighter” ($x_1$) and “redder-dimmer” ($c$) trends motivate the luminosity correlations.

Calibrators are SNeIa from host galaxies whose distances have been cross-calibrated with Cepheids. Calibrator distances serve as host-galaxy distance anchors \cite{2022ApJ...934L...7R}; they are included in the likelihood to mitigate a degeneracy in $H_0$ and $M$, whose impact on cosmological constraints cannot be differentiated using supernovae alone \cite{2022ApJ...938..110B}.

Predicted distances are redshift- and cosmology-dependent, i.e.,
\begin{equation}
    \mu_{\mathrm{model}}(z_i)=5\log(d_L(z_i)/10\,\mathrm{pc}),
\end{equation}
where 
\begin{equation}\label{eqn:luminosity_distance}
    d_L(z)=(1+z)\frac{c}{H_0}\int_0^z\frac{dz'}{E(z')}
\end{equation}
is the luminosity distance and  
\begin{equation}\label{eqn:hubble_parameter}
        E(z)=\sqrt{\Omega_M(1+z)^3+\Omega_k(1+z)^2+\Omega_\lambda(1+z)^{3(1+w)}},
\end{equation}
is the normalized redshift-dependent expansion rate \cite{1993ppc..book.....P,1999astro.ph..5116H}. Four (4) cosmological models are considered in this work:
\begin{itemize}
    \item Flat-$\Lambda$CDM: Free parameters $M$, $H_0$, $\Omega_M$; fixed parameters $w=-1$, $\Omega_k=0$. 
    \item $\Lambda$CDM: Free parameters $M$, $H_0$, $\Omega_M$ and $\Omega_\Lambda$; fixed parameter $w=-1$.
    \item Flat-$w$CDM: Free parameters  $M$, $H_0$, $\Omega_M$ and $w$; fixed parameter $\Omega_k=0$.
    \item Flat-$w_\circ w_a$CDM\cite{2001IJMPD..10..213C,2003PhRvL..90i1301L}: $w=w_\circ+w_a z/(1+z)$, free parameters $M$, $H_0$, $\Omega_M$, $w_\circ$, and $w_a$; fixed parameter $\Omega_k=0$,
\end{itemize}
where $\Omega_M$, $\Omega_k$, and $\Omega_\Lambda$ are dimensionless parameters that describe the mass, spatial curvature, and dark energy content of the Universe ($\Omega_M+\Omega_k+\Omega_\Lambda=1$), and $w$ specifies the dark energy equation of state. 

\begin{table}
\caption{\label{tab:pantheon_subsets}Pantheon+ subset characteristics summary. Subsets were defined based on homogeneity, i.e., the fraction of SNeIa with $|\alpha|\leq\alpha_\mathrm{max}$.}
\begin{ruledtabular}
\begin{tabular}{ccccc}
Homogeneity & No. SNe & No. Distance & $\alpha_\mathrm{max}$ & $\sigma^2_\alpha$ \\ 
 Subset & (\% of Total) & Calibrators & & \\ \hline 
A & 34 (2\%) & 1 & 0.025 & $1.9\times10^{-4}$ \\ 
B & 85 (5\%) & 4 & 0.062 & $1.3\times10^{-3}$ \\ 
C & 170 (10\%) & 7 & 0.12 & $5.4\times10^{-3}$ \\ 
D & 340 (20\%) & 13 & 0.23 & $1.9\times10^{-2}$ \\ 
E & 850 (50\%) & 27 & 0.61 & $0.11$ \\ 
F & 1701 (100\%) & 43 & 3.81 & $0.97$  
\end{tabular}
\end{ruledtabular}
\end{table}

\begin{table*}
\caption{\label{tab:pantheon_subset_results}Pantheon+ subset analysis summary. Likelihoods for each subset were independently minimized to obtain the tabulated cosmological parameter constraints. Subsets defined based on homogeneity (Tab.~\ref{tab:pantheon_subsets}).}
\begin{ruledtabular}
\begin{tabular}{ccccccccc}
Homogeneity & Model & $M$ & $H_0$ & $\Omega_m$ & $\Omega_{\Lambda}$ & $w$ & $w_a$ \\ 
 Subset & & (mag) & (km s$^{-1}$ Mpc$^{-1}$) & & & & \\ \hline \\
A & Flat-$\Lambda$CDM & $-19.33^{+0.08}_{-0.10}$ & $69.2^{+3.2}_{-3.1}$ & $0.351^{+0.069}_{-0.092}$ & $0.649^{+0.091}_{-0.069}$ & & \\
& $\Lambda$CDM & $-19.34\pm0.09$ & $69.2^{+3.1}_{-3.7}$ & $0.324^{+0.136}_{-0.125}$ & $0.669^{+0.220}_{-0.298}$ & & \\
& Flat-$w$CDM & $-19.34^{+0.09}_{-0.10}$ & $69.2^{+3.5}_{-3.2}$ & $0.448^{+0.080}_{-0.194}$ & $0.552^{+0.194}_{-0.080}$ & $-0.959^{+0.336}_{-0.544}$ & \\
& Flat-$w_\circ w_a$CDM & $-19.34^{+0.09}_{-0.10}$ & $68.6^{+4.0}_{-2.9}$ & $0.475^{+0.079}_{-0.147}$ & $0.525^{+0.147}_{-0.079}$ & $-1.127^{+0.516}_{-0.508}$ & $0.044^{+0.686}_{-3.376}$ \\ \\
B & Flat-$\Lambda$CDM & $-19.33^{+0.05}_{-0.07}$ & $69.0^{+2.4}_{-2.0}$ & $0.387^{+0.053}_{-0.080}$ & $0.613^{+0.078}_{-0.053}$ & & \\
& $\Lambda$CDM & $-19.33\pm0.06$ & $69.3^{+2.2}_{-2.0}$ & $0.365^{+0.139}_{-0.117}$ & $0.655^{+0.211}_{-0.220}$ & & \\
& Flat-$w$CDM & $-19.34^{+0.07}_{-0.05}$ & $69.1^{+2.6}_{-1.8}$ & $0.478^{+0.081}_{-0.123}$ & $0.522^{+0.121}_{-0.081}$ & $-1.411^{+0.557}_{-0.377}$ & \\
& Flat-$w_\circ w_a$CDM & $-19.33^{+0.06}_{-0.07}$ & $69.2^{+2.5}_{-2.1}$ & $0.512^{+0.067}_{-0.094}$ & $0.488^{+0.094}_{-0.067}$ & $-1.387^{+0.336}_{-0.528}$ & $0.191^{+0.862}_{-3.477}$ \\ \\
C & Flat-$\Lambda$CDM & $-19.28\pm0.05$ & $71.7\pm1.7$ & $0.373^{+0.041}_{-0.059}$ & $0.627^{+0.059}_{-0.042}$ & & \\
& $\Lambda$CDM & $-19.28^{+0.04}_{-0.06}$ & $71.9^{+1.4}_{-1.9}$ & $0.363^{+0.117}_{-0.119}$ & $0.610^{+0.198}_{-0.149}$ & & \\
& Flat-$w$CDM & $-19.29^{+0.05}_{-0.04}$ & $71.9\pm1.8$ & $0.486^{+0.064}_{-0.144}$ & $0.514^{+0.146}_{-0.064}$ & $-1.005^{+0.196}_{-0.664}$  & \\
& Flat-$w_\circ w_a$CDM & $-19.28\pm0.05$ & $72.0^{+1.7}_{-1.9}$ & $0.497^{+0.072}_{-0.088}$ & $0.503^{+0.088}_{-0.072}$ & $-1.203^{+0.288}_{-0.604}$ & $0.156^{+0.914}_{-2.702}$ \\ \\
D & Flat-$\Lambda$CDM & $-19.28^{+0.05}_{-0.04}$ & $72.5^{+1.6}_{-1.3}$ & $0.309^{+0.040}_{-0.029}$ & $0.691^{+0.029}_{-0.042}$ & & \\ 
& $\Lambda$CDM & $-19.28\pm0.04$ & $72.8^{+1.2}_{-1.9}$ & $0.323^{+0.086}_{-0.101}$ & $0.664^{+0.141}_{-0.127}$ & & \\
& Flat-$w$CDM & $-19.28^{+0.05}_{-0.04}$ & $72.8^{+1.5}_{-1.7}$ & $0.398^{+0.082}_{-0.110}$ & $0.602^{+0.112}_{-0.082}$ & $-1.050^{+0.284}_{-0.413}$  & \\ 
& Flat-$w_\circ w_a$CDM & $-19.27^{+0.04}_{-0.05}$ & $73.2^{+1.2}_{-1.9}$ & $0.454^{+0.067}_{-0.115}$ & $0.546^{+0.115}_{-0.067}$ & $-1.095^{+0.280}_{-0.444}$ & $0.303^{+0.964}_{-2.204}$ \\ \\
E & Flat-$\Lambda$CDM & $-19.23\pm0.03$ & $74.0^{+0.9}_{-1.2}$ & $0.335^{+0.019}_{-0.026}$ & $0.665^{+0.026}_{-0.019}$ & & \\
& $\Lambda$CDM & $-19.24\pm0.03$ & $73.4^{+1.5}_{-0.7}$ & $0.260^{+0.093}_{-0.046}$ & $0.571^{+0.128}_{-0.079}$ & & \\
& Flat-$w$CDM & $-19.23\pm0.03$ & $74.0^{+0.9}_{-1.3}$ & $0.304^{+0.088}_{-0.083}$ & $0.696^{+0.082}_{-0.088}$ & $-0.886^{+0.195}_{-0.187}$ & \\
& Flat-$w_\circ w_a$CDM & $-19.24^{+0.04}_{-0.03}$ & $73.9^{+1.0}_{-1.4}$ & $0.376^{+0.090}_{-0.106}$ & $0.624^{+0.106}_{-0.090}$ & $-0.866^{+0.148}_{-0.224}$ & $0.197^{+0.719}_{-1.548}$ \\ \\
F & Flat-$\Lambda$CDM & $-19.25^{+0.03}_{-0.02}$ & $73.9^{+0.7}_{-1.3}$ & $0.331^{+0.019}_{-0.016}$ & $0.669^{+0.016}_{-0.019}$ & & \\
& $\Lambda$CDM & $-19.24^{+0.04}_{-0.02}$ & $73.3^{+1.3}_{-0.7}$ & $0.304^{+0.056}_{-0.050}$ & $0.629^{+0.078}_{-0.082}$ & & \\
& Flat-$w$CDM & $-19.25\pm0.03$ & $73.7^{+0.7}_{-1.4}$ & $0.302^{+0.070}_{-0.061}$ & $0.698^{+0.059}_{-0.070}$ & $-0.891^{+0.120}_{-0.171}$ & \\
& Flat-$w_\circ w_a$CDM & $-19.25\pm0.03$ & $73.3^{+1.0}_{-1.1}$ & $0.401^{+0.065}_{-0.085}$ & $0.599^{+0.083}_{-0.065}$ & $-0.871^{+0.088}_{-0.192}$ & $-0.308^{+1.026}_{-1.970}$ \\ \\
\end{tabular}
\end{ruledtabular}
\end{table*}

\section{\label{sec:impact}The Impact of Diversity}

Six subsets were constructed using the full Pantheon+ dataset (Table~\ref{tab:pantheon_subsets}): Subsets-A through -F correspond to supernova ensembles with the smallest 2, 5, 10, 20, 50, and 100\% of $\lvert\alpha\rvert$ values, respectively. The subsets are distinguished by their sample size and degree of homogeneity, i.e., the dispersion of $\alpha$-values within each subset ($\sigma_\alpha^2$).

A likelihood (Eqn.~\ref{eqn:likelihood}) was constructed for each subset by extracting the matrix elements of its members from the archived full-sample Pantheon+ likelihood without modification. Each likelihood was minimized using the PolyChord sampler \cite{2015MNRAS.453.4384H} in the CosmoSIS package \cite{2015A&C....12...45Z} with 250 live points, 30 repeats, and an evidence tolerance of 0.1.  

\begin{figure}[!]
    \includegraphics[scale=0.33]{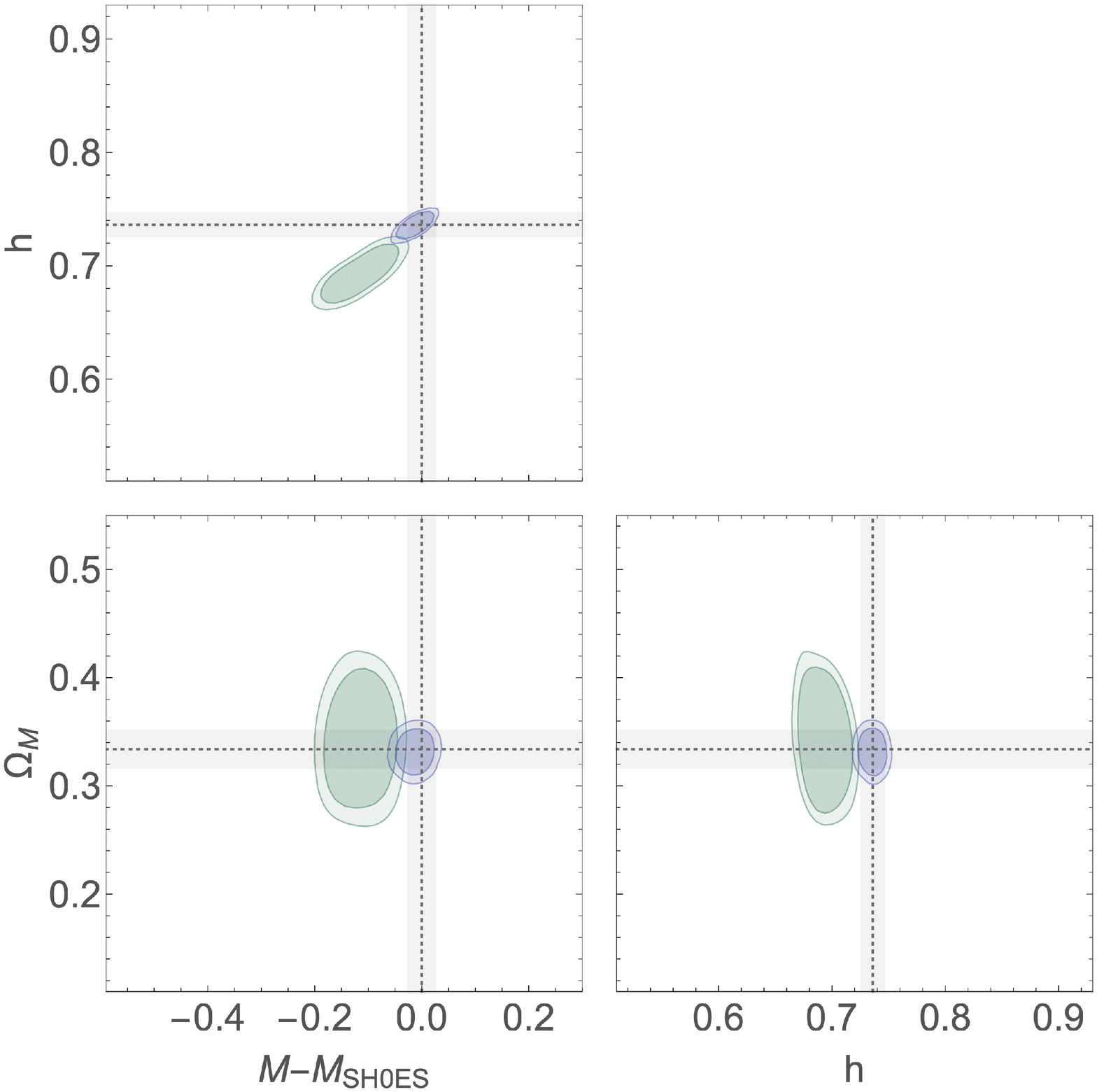}
    \caption{\label{fig:FlatLambdaCDM-constraints}Constraints on Flat-$\Lambda$CDM model parameters ($M-M_{\mathrm{SH0ES}}$, $h$, $\Omega_M$) from Pantheon+. Shown here are the $68\%$ and $95\%$ confidence contours obtained independently using Subset-A (green) and Subset-F (blue). Subset-A is the most homogeneous supernova sample; Subset-F corresponds to the full Pantheon+ sample and is the most diverse. Previously reported Pantheon+ confidence limits are also shown (gray bands) \cite{2022ApJ...938..110B}. The Hubble-Lemaitre parameter is given by $100\,h$ km s$^{-1}$ Mpc$^{-1}$.} 
\end{figure}

\begin{figure*}[!]
    \includegraphics[scale=0.33]{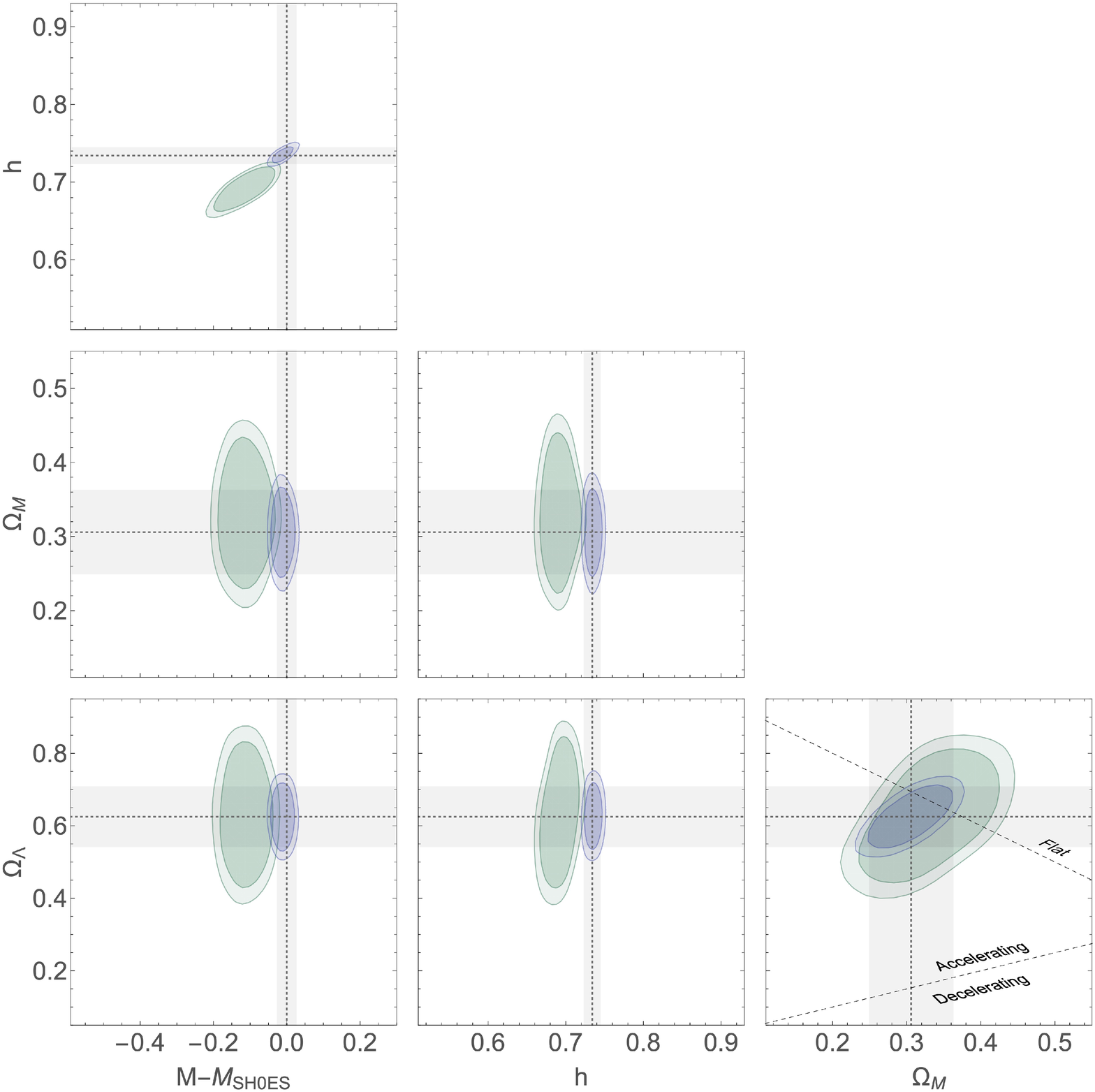}
    \caption{\label{fig:LambdaCDM-constraints}Same as Fig.~\ref{fig:FlatLambdaCDM-constraints} for a $\Lambda$CDM cosmology ($M-M_{\mathrm{SH0ES}}$, $h$, $\Omega_M$, $\Omega_\Lambda$). Two dashed lines are shown for reference: a flat universe where $\Omega_M+\Omega_\Lambda=1$ and another indicative of an accelerating universe.} 
\end{figure*}

\begin{figure*}[!]
    \includegraphics[scale=0.33]{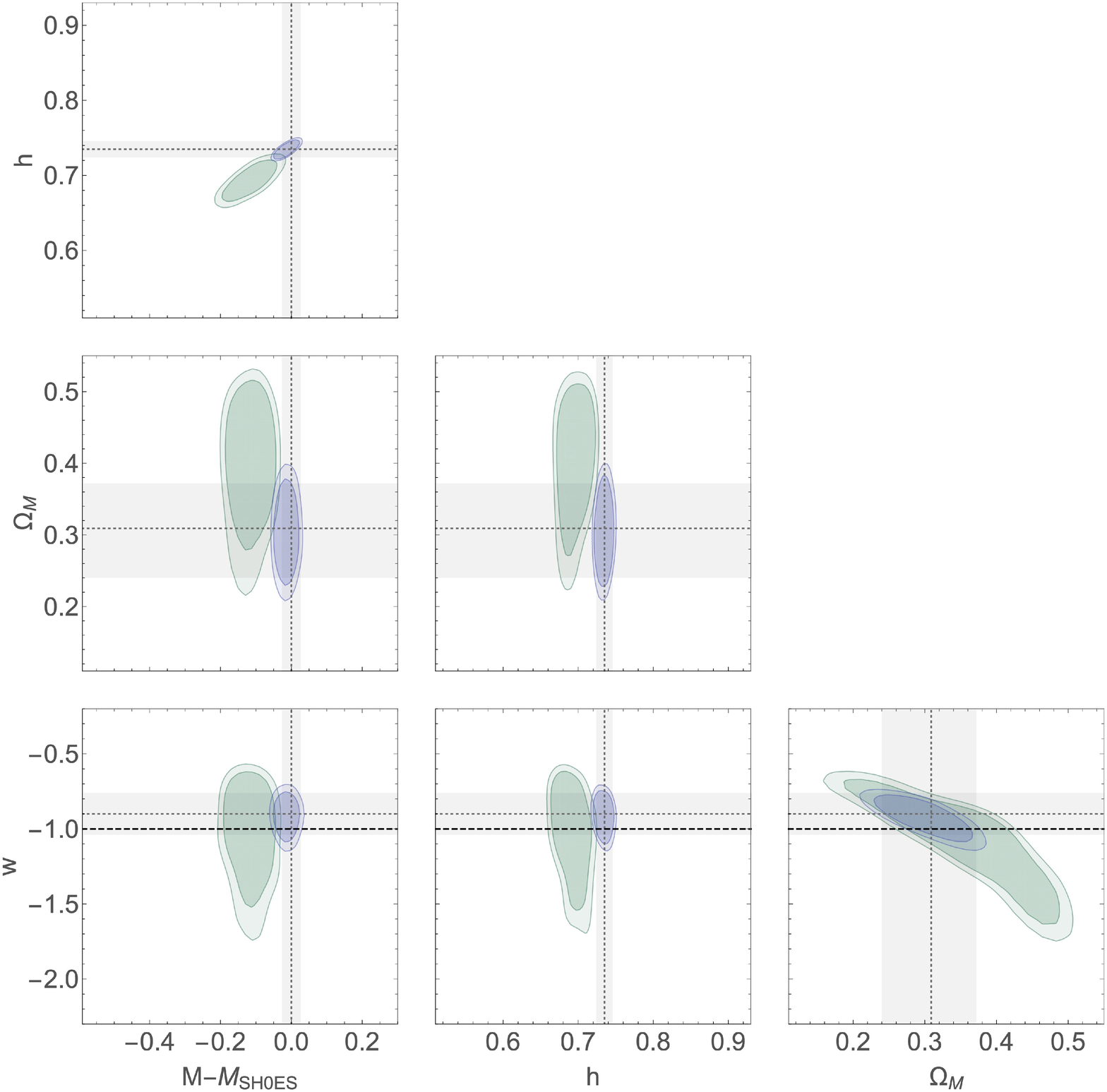}
    \caption{\label{fig:FlatwCDM-constraints}Same as Fig.~\ref{fig:FlatLambdaCDM-constraints} for a Flat-wCDM cosmology ($M-M_{\mathrm{SH0ES}}$, $h$, $\Omega_M$, $w$). The equation of state parameter consistent with a cosmological constant is also shown ($w=-1$; dashed).} 
\end{figure*}

\begin{figure*}[!]
    \includegraphics[scale=0.33]{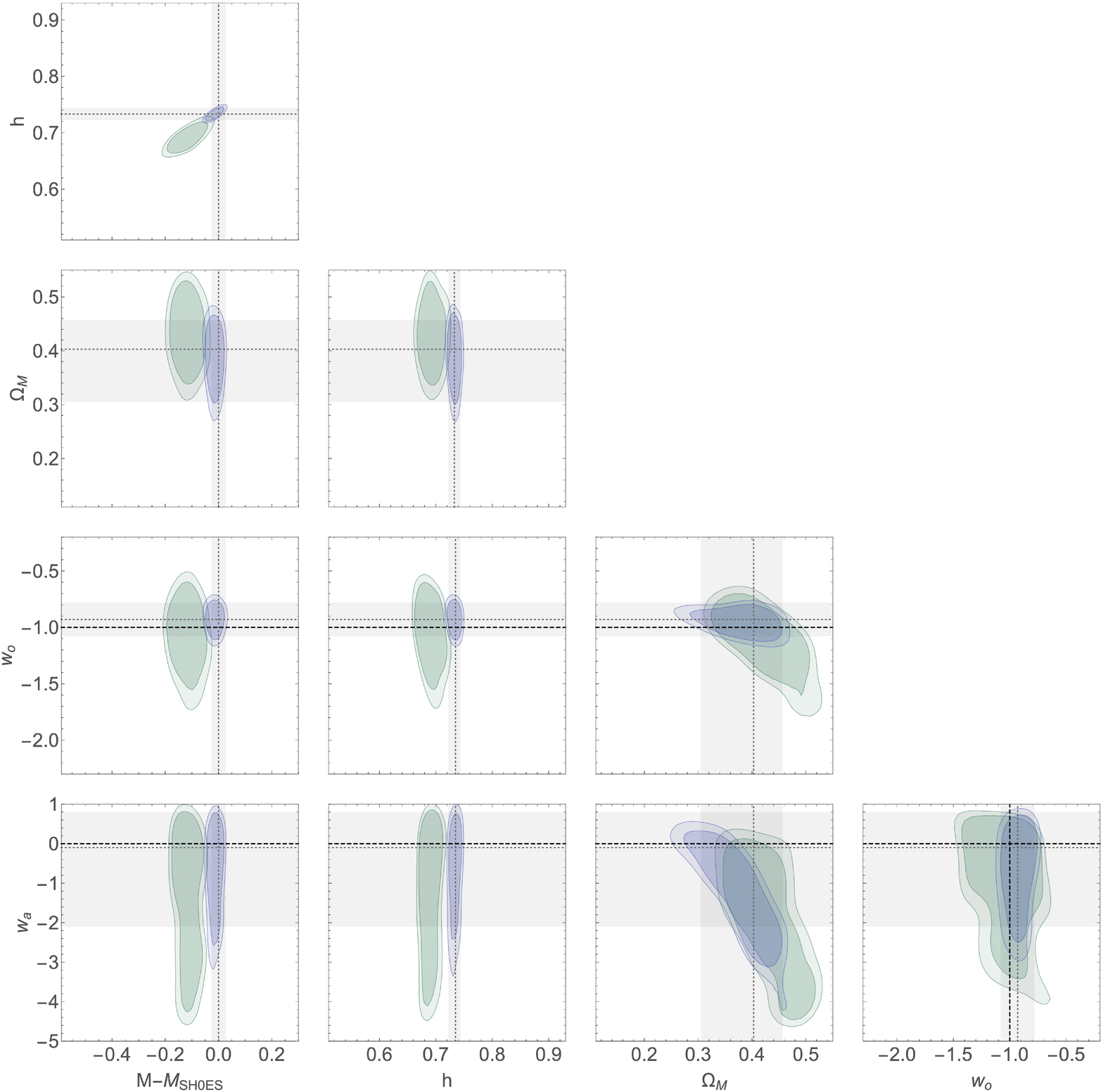}
    \caption{\label{fig:Flatw0waCDM-constraints}Same as Fig.~\ref{fig:FlatLambdaCDM-constraints} for a Flat-$w_\circ w_a$CDM cosmology ($M-M_{\mathrm{SH0ES}}$, $h$, $\Omega_M$, $w_\circ$, $w_a$). Equation of state parameters consistent with a cosmological constant are also shown ($w_\circ=-1$, $w_a=0$; dashed).} 
\end{figure*}

Parameter constraints for each subset are given in Table~\ref{tab:pantheon_subset_results}. Results from Subset-F are commensurate with the constraints reported by Pantheon+ \cite{2022ApJ...938..110B}. The constraints on $M$ and $H_0$ from Subset-A (least diverse) and Subset-F (most diverse) are incompatible at $\sim$$95\%$ confidence level (C.L.), independent of cosmological model (Fig.~\ref{fig:FlatLambdaCDM-constraints}-\ref{fig:Flatw0waCDM-constraints}). 

\begin{figure}[!]
    \includegraphics[scale=0.40]{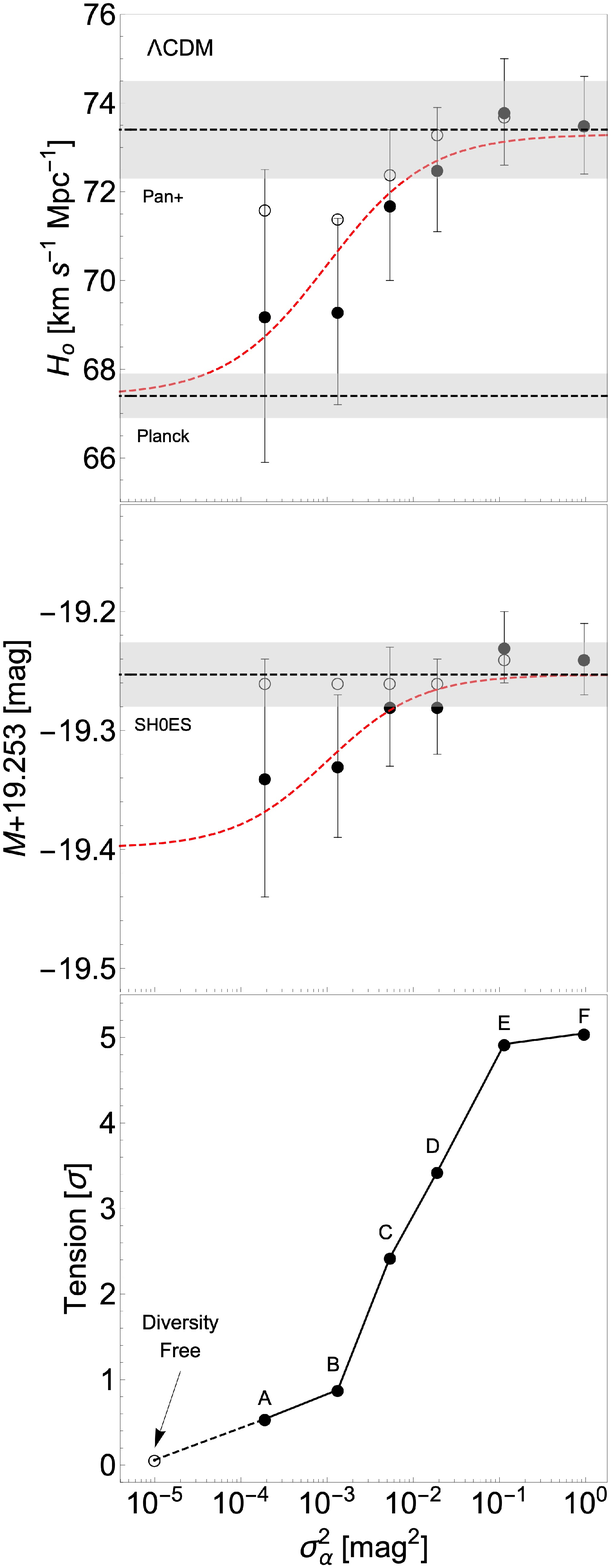}
    \caption{\label{fig:hubble_trends}Parameter trends for a $\Lambda$CDM cosmology. Constraints on $H_0$ (top), $M$ (center), and the Hubble Tension (bottom) versus the degree of subset diversity, $\sigma_\alpha^2$. Closed circles ($\bullet$) denote constraints obtained using the subsets defined in Table~\ref{tab:pantheon_subsets}; open circles ($\circ$) are either constraints that include all calibrators (top and middle; error bars removed for clarity) or the Tension in the diversity-free limit (bottom). Constraints from Planck \cite{2020A&A...641A...6P}, Pantheon+ \cite{2022ApJ...938..110B}, and SH0ES \cite{2022ApJ...934L...7R} are also shown. The fits (red) and Tension are given by Equations~\ref{eqn:sigmoid} and \ref{eqn:hubble_tension_defn}, respectively.} 
\end{figure}

Both $M$ and $H_0$ show a subset (diversity) dependence that other parameters don't. The subset-dependent trends were parameterized using a sigmoid logistic function of the form 
\begin{equation}\label{eqn:sigmoid}
    p(x)=p_\mathrm{SNe}+\frac{p_\mathrm{F}-p_\mathrm{SNe}}{(1+\mathrm{Exp}[{-v(x+3}])},
\end{equation}
where $p$ is the parameter of interest, $x=\log_{10}{\sigma_\alpha^2}$, $v$ is the rate of change between the two endpoints at $x=(-6,0)$, $p_\mathrm{F}$ is the Subset-F parameter estimate, and $p_\mathrm{SNe}$ is a free parameter. Representative fits are shown in Fig.~\ref{fig:hubble_trends} for a $\Lambda$CDM cosmology; similar results hold for all cosmological models. 

Parameter constraints from SH0ES and Pantheon+ are inconsistent with the fitted trends. A subset-independent $M$, equal to the SH0ES estimate \cite{2022ApJ...934L...7R}, is excluded at $\sim$$80\%$ C.L. ($\Delta\chi^2=3.3$); Pantheon+ determinations of $H_0$ \cite{2022ApJ...938..110B} are excluded at $\geq$$95\%$ C.L. ($\Delta\chi^2=6.1$). Similar results hold for all cosmological models. 

Additional subset analyses confirm/reveal:
\begin{itemize}
    \item Observation: The SH0ES estimate of $M$ is recovered when all calibrators are included in subset likelihoods, independent of their $\alpha$ values (e.g., Fig.~\ref{fig:hubble_trends}, top, open circles). Impact: Constraints on $M$ depend solely on calibrator supernovae (Sec.~\ref{sec:cosmology}).
    \item Observation: Constraints on $M$ skew to larger magnitudes as more diverse SH0ES calibrators are included in the likelihoods (e.g., Fig.~\ref{fig:hubble_trends}, top, solid circles). Impact: Calibrator diversity (Fig.~\ref{fig:sys_vs_redshift}) influences constraints on $M$.  
    \item Observation: Constraints on $M$ and $H_0$ are highly correlated (Pearson $r$=0.984), a reflection of their intrinsic degeneracy (Sec.~\ref{sec:cosmology}). Impact: Diversity-driven overestimates in $M$ bias $H_0$ constraints to artificially large values. 
    \item Observation: Pantheon+ determinations of $H_0$ are not recovered when all calibrators are included (e.g., Fig.~\ref{fig:hubble_trends}, top); they remain excluded at $\sim$$68\%$ C.L. Impact: Calibrator-driven bias in $H_0$ is exacerbated by the degree of diversity within Pantheon+ overall.  
\end{itemize}
The observed trends are evidence of diversity-driven systematic effects that bias Pantheon+-derived $H_0$ constraints to large values.

\subsection{Validation of Diversity Mitigation}

Diversity-driven systematic effects manifest in the peak magnitudes used to infer supernovae distances (Eqn.~\ref{eqn:tripp}). Event selections based on a relative measure of diversity ($\alpha$) were required to quantify the impact of these systematics (Sec.~\ref{sec:impact}). The efficacy of absolute corrections to directly mitigate diversity systematics was tested using supernova siblings. 

Sibling distance moduli are known to be unequal \cite{2020ApJ...896L..13S,2022ApJ...938..113S}, an observation at odds with their definition. Paired differences, $\Delta\mu$=$\lvert\mu_1-\mu_2\rvert$, should be zero, yet all sibling datasets are inconsistent with this requirement (Table~\ref{tab:siblings_table}). For example, the mean difference between Pantheon+ natural sibling pairs is $\Delta\mu$=$0.212\pm0.035$, a 4.6$\sigma$ deviation from zero ($t$-statistic=5.98). In contrast, the mean of modified differences, $\Delta\mu-\alpha_\mathrm{sib}$=$-0.030\pm0.044$, differs from zero by $<$1$\sigma$ ($t$-statistic=0.70). Similar changes were observed for each dataset (Table~\ref{tab:siblings_table}).

The reduction in sibling distance differences is evidence that shape-dependent peak magnitude adjustments mitigate diversity-driven systematic effects. Sibling-based tests are an important confirmation that relative $\alpha$-based event selections minimize peak magnitude dispersion, and hence the distance moduli that define the SNeIa-defined Hubble Diagram.

\section{\label{sec:tension}Easing the Tension}

\begin{figure*}
    \includegraphics[scale=0.5]{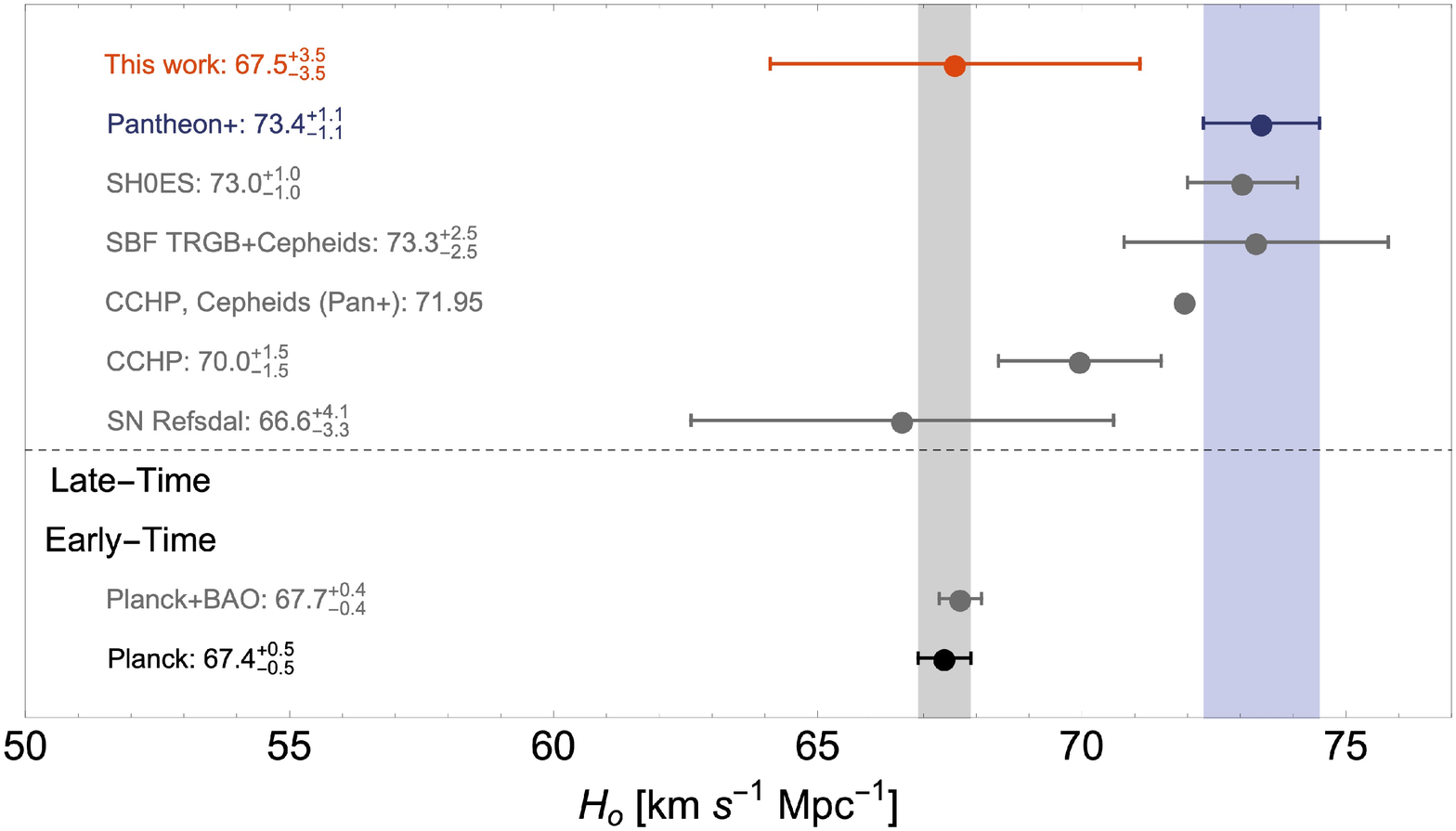}
    \caption{\label{fig:hubble_comparison}Comparison of recent estimates Hubble-Lema\^{i}tre parameter ($H_0$) estimates. Constraints extracted from the Pantheon+ diversity analyses are shown (red) along with other early- and late-time determinations. Early-time constraints include the baseline Planck and the joint Planck + Baryon Acoustic Oscillation (BAO) CMB analyses for a $\Lambda$CDM cosmology \cite{2020A&A...641A...6P}; late-time constraints include the original Pantheon+ result \cite{2022ApJ...938..110B}, SH0ES \cite{2022ApJ...934L...7R}, surface brightness fluctuations (SBF) TRGB+Cepheids \cite{2021ApJ...911...65B}, the Carnegie-Chicago Hubble Program (CCHP) adopted and Pantheon+ re-analysis values \cite{freedman2024statusreportchicagocarnegiehubble}, and the preferred subset of gravitational lensing models for SN Refsdal \cite{2023Sci...380.1322K}. The gray vertical band corresponds to the baseline Planck estimate ($H_0=67.4\pm0.5$ km s$^{-1}$ Mpc$^{-1}$), the blue band to the previously published Pantheon+ value ($H_0=73.4\pm1.1$ km s$^{-1}$ Mpc$^{-1}$).} 
\end{figure*}

A "diversity free" determination of the Hubble-Lema\^{i}tre parameter was obtained by extrapolating Equation~\ref{eqn:sigmoid} to the low $\alpha$-dispersion regime: $H_0$=$67.5\pm3.5$ km s$^{-1}$ Mpc$^{-1}$ (68\% C.L.) at $\sigma_\alpha^2$$=$$10^{-5}$. Other recent estimates of $H_0$ are shown in Fig.~\ref{fig:hubble_comparison}. 

Tension between early- and late-universe estimates of $H_0$ is given by 
\begin{equation}\label{eqn:hubble_tension_defn}
  \mathrm{Tension}=\frac{H_{0,\mathrm{Subset}}-H_{0,\mathrm{Planck}}}{\sqrt{\sigma_{H_0,\mathrm{Subset}}^2+\sigma_{H_0,\mathrm{Planck}}^2}},
\end{equation}
where $H_{0,\mathrm{Subset}}$ is a Pantheon+-derived constraint, and $H_{0,\mathrm{Planck}}$ is the Planck baseline ($H_0=67.4\pm0.5$ km s$^{-1}$ Mpc$^{-1}$) \cite{2020A&A...641A...6P}). The subset-dependent Hubble Tension is shown in Fig.~\ref{fig:hubble_trends} (bottom), along with the diversity-free Tension. Similar trends hold for all cosmological models. 

Mitigation of diversity-driven systematic effects eases the Tension from $\sim$$5\sigma$ for the full Pantheon+ sample to $0.06\sigma$ in the "diversity free" limit, a 98\% reduction. Easing of the Tension is due to a convergence of the $H_0$ constraint with the Planck determination; it cannot be explained by parameter uncertainties alone.

\subsection{Discussion}

Diversity-driven systematic effects dominate previously reported SNeIa-derived constraints on $H_0$. The total diversity-driven systematic bias in $H_0$ is $\Delta H_0$=$5.8\pm2.5$ km s$^{-1}$ Mpc$^{-1}$ ($\Lambda$CDM), the difference between diversity free and Subset-F determinations. Calibrator diversity accounts for $\Delta H_0^\mathrm{cal}$=$2.8\pm1.1$ km s$^{-1}$ Mpc$^{-1}$ of the total, or $48.3\pm1.8\%$. The total bias is $\sim$8$\times$ larger than the combined statistical and systematic uncertainties described by Pantheon+, and $\sim$20$\times$ larger than the systematic uncertainty alone \cite{2022ApJ...938..110B}.

Pantheon+ Hubble diagram residuals have also been underestimated, the direct result of previous $H_0$ constraints biased to a large value by intrinsic SNeIa diversity. The total observed scatter (rms) is 0.23 mag for the best fit Flat-$\Lambda$CDM cosmology, more than 50\% larger than the Pantheon+ estimate of 0.15 mag ($z$$>$0.01) \cite{2022ApJ...938..110B}; similar results hold for all cosmological models.

Additional circumstantial evidence for intrinsic thermonuclear diversity exists but has been attributed to other causes. The Chicago-Carnegie Hubble Program's (CCHP) reanalysis of Pantheon+ is an illustrative example. That effort used Cepheids in the host galaxies of 11 SNeIa to re-calibrate the extragalactic distance scale. The difference between CCHP and Subset-F determinations of $H_0$, $\Delta H_0^\mathrm{CCHP}$=1.35 km s$^{-1}$ Mpc$^{-1}$ \cite{freedman2024statusreportchicagocarnegiehubble}, is governed solely by calibrator effects. 

The CCHP calibrators are $\delta$=$52.2\%$ less diverse than the 42 SH0ES calibrators used by Pantheon+ ($\alpha$=$0.117$ mag versus $\alpha$=$0.245$ mag, on average). An independent assessment of calibrator diversity, $\delta$=$51.8\pm18.9\%$, was obtained using $\Delta H_0^\mathrm{CCHP}$=$\Delta H_0^\mathrm{cal}\times(1-\delta)$, which approximates the reduction in $H_0$ due to calibrator diversity. Agreement with the CCHP calibrators supports the conclusion that serendipitous diversity mitigation, rather than re-calibration of the distance scale \cite{2019ApJ...882...34F,2020ApJ...891...57F,2021ApJ...919...16F,freedman2024statusreportchicagocarnegiehubble}, is responsible for easing of the Hubble Tension.

\section{\label{sec:summary}Summary}

Accuracy gains in supernova cosmology demand a recognition that thermonuclear diversity complicates the notion of SNeIa as standardizable candles. The diversity-driven systematic effects quantified here are evidence of intrinsic differences between SNeIa. Diverse supernovae violate the standard candle premise, alter the supernova-derived Hubble Diagram, and skew cosmological parameter constraints. 

Late-universe estimates of the Hubble-Lema\^{i}tre parameter ($H_0$) have been biased to artificially large values by SNeIa diversity. Previous investigations, like Pantheon+, have not accounted for the intimate relationship between (standardized) peak magnitudes and light curve shape, the latter a proxy for the diverse thermonuclear scenarios that govern SNeIa. Event selection based on relative differences in shape-dependent peak magnitudes were used to quantify, and mitigate, the impact of diversity. 

Diversity-driven systematic effects fully explain the Hubble Tension, a conclusion that contrasts pointedly with \citet{2022ApJ...938..110B}. This agreement weakens the need to invoke new physics in response to the significant disparity in cosmic expansion rate estimates reported over the past decade. All cosmological parameters considered here ($H_0$, $\Omega_m$, $\Omega_{\Lambda}$, $w$) are consistent with a flat $\Lambda$CDM cosmology; they are also in concordance with early-Universe constraints after accounting for the impact of diversity.

Supernova statistics alone are insufficient to address the impact of diversity on future supernova cosmology investigations. An alternative path forward, based on the techniques presented here, leverages the phenomenology of SNeIa light curves. Large numbers of detected supernovae are required to compensate for the additional diversity-mitigating event selection criteria: $\leq$1\% of the Pantheon+ sample survives to "diversity free" dispersion levels ($\sigma_\alpha^2$$\leq$$10^{-5}$), too few to facilitate a meaningful likelihood analysis. 

The Vera C. Rubin Observatory is well suited to support diversity-mitigated supernova cosmology: more than $10^4$ SNeIa yr$^{-1}$ will be detected and monitored as part of Rubin's main survey \cite{lsstsciencecollaboration2009lsstsciencebookversion}. Approximately 100 SNeIa yr$^{-1}$ will satisfy the selection criteria if the diversity of Rubin supernovae is similar to Pantheon+. An ensemble of Rubin-detected standard candles will have a degree of diversity $\sim$$10^{-5}$ times smaller than Pantheon+ to provide both accurate and precise cosmological model constraints. A 1\% determination of $H_0$ is expected in $\sim$3-5 years of operation.

\begin{acknowledgments}
This work was supported in part by NASA grants 80NSSC22K0623 (ADAP) and 80NSSC24K0769 (ATP). The author thanks Drs. Patrick Peplowski, Jack Wilson, Robert Schaefer, John Beacom, Chris Fryer, and Breanna Crane for useful feedback and conversations. The author also acknowledges the support of Drs. Scott Murchie, Louise Procktor, and Jason Kalirai.
\end{acknowledgments}

\bibliography{SiblingRivalry}% Produces the bibliography via BibTeX.
\bibliographystyle{unsrtnat}

\end{document}